\documentclass[12pt]{iopart}

\usepackage{graphicx} 
\usepackage{subcaption}
\usepackage{ulem}
\usepackage{hyperref}
  \hypersetup{colorlinks,linkcolor=[rgb]{0.2,0.2,0.75},citecolor=[rgb]{0.3,0.5,0.5},urlcolor=[rgb]{0.4,0.4,0.6},breaklinks}
\usepackage{doi}

\usepackage[T1]{fontenc}
\usepackage[utf8]{inputenc}
\usepackage{xcolor}

\usepackage{algorithm}
\usepackage{algpseudocode}

\usepackage[numbers,sort&compress]{natbib}
\begin{document}

\title[Matrix compression of response matrices in JOREK]{Implementation of matrix compression in the coupling of JOREK to realistic 3D conducting wall structures}

\author{F.~Cipolletta$^1$, N.~Schwarz$^2$, M.~Hoelzl$^2$, S.~Ventre$^3$, N.~Isernia$^4$, G.~Rubinacci$^5$, A.~Soba$^1$, M.~J.~Mantsinen$^{1,6}$, and the JOREK~Team$^7$}

\date{October 2023}

\address{$^1$ Barcelona Supercomputing Center (BSC), Barcelona, Spain}
\address{$^2$ Max Planck Institute for Plasma Physics, Boltzmannstr. 2, 85748 Garching b. M., Germany}
\address{$^3$ Dipartimento di Ingegneria Elettrica e dell’Informazione, Università degli Studi di Cassino e del Lazio Meridionale, Via Gaetano di Biasio 43, 03043 Cassino, Italy}
\address{$^4$ Dipartimento di Ingegneria Elettrica e delle Tecnologie dell'Informazione, Università degli Studi di Napoli Federico II, via Claudio 21, 80125 Napoli, Italy}
\address{$^5$ CREATE Consortium, Università degli Studi di Napoli Federico II, via Claudio 21, 80125 Napoli, Italy}
\address{$^6$ ICREA, Barcelona, Spain}
\address{$^7$ See the author list of ref.~\cite{hoelzl2021jorek}}

\ead{federico.cipolletta@bsc.es}
\vspace{10pt}
\begin{indented}
\item[]\today
\end{indented}


\begin{abstract}
JOREK is an advanced non--linear simulation code for studying MHD instabilities in magnetically confined fusion plasmas and their control and/or mitigation. A free--boundary and resistive wall extension was introduced via coupling to the STARWALL and CARIDDI codes, both able to provide dense response matrices describing the electromagnetic interactions between plasma and conducting structures. For detailed CAD representations of the conducting structures and high resolutions for the plasma region, memory and computing time limitations restrict the possibility of simulating the ITER tokamak. In the present work, the Singular Value Decomposition provided by routines from the ScaLAPACK library has been successfully applied to compress some of the dense response matrices and thus optimize memory usage. This is demonstrated for simulations of Tearing Mode and Vertical Displacement Event instabilities. An outlook to future applications on large production cases and further extensions of the method are discussed.
\end{abstract}


\submitto{\PPCF}

\maketitle

\section{Introduction}
\label{sec:Intro}

The ability to perform realistic predictive simulations of large-scale plasma instabilities in future magnetic confinement devices like ITER or DEMO is essential for their safe and efficient operation. Towards this effort, it is necessary to develop powerful multi-physics models, validate them against experiments in existing fusion devices, and optimize the models algorithmically and numerically for predictive simulations with particularly high resolutions. Only after reaching the capability to produce robust simulations of large device sizes subject to challenging plasma conditions, actual predictions can be made with sufficiently high fidelity. The present article describes efforts aiming to enable simulations that were previously not feasible, by implementing and verifying matrix compression techniques in the coupling of the JOREK non-linear MHD code~\cite{czarny2008bezier,huysmans2007mhd,hoelzl2021jorek} with the resistive wall codes STARWALL~\cite{merkel2007feedback,Merkel_SW} and CARIDDI~\cite{albanese1988integral,albanese1997finite}. 

Accounting for resistive wall effects is indispensable for capturing magnetic field perturbations across the whole domain with full realism. Most notably, certain dangerous classes of instabilities like vertical displacement events (VDEs)~\cite{artola2023modelling} or resistive wall modes (RWMs)~\cite{igochine2012physics} would not be captured at all without these contributions. For recent applications of the JOREK code including instabilities at the plasma boundary and their control, major disruptions and their mitigation by shattered pellet injection, runaway electrons, VDEs, divertor and scrape-off layer dynamics, turbulence, and more, the reader is referred to the review article Ref.~\cite{hoelzl2024jorek} and references therein. In the rest of this section, we briefly describe the JOREK code, its coupling to the resistive wall codes, and the aim of the work described in the rest of this article.

JOREK is a non-linear code for global simulations of instabilities in the realistic geometry of magnetic confinement fusion devices. The available physics models include reduced and full MHD fluid descriptions of the plasma with various extensions for two-fluid effects, neoclassical physics, kinetic treatment for specific particle species, etc., a variety of hybrid kinetic-fluid models, and (gyro)kinetic turbulence models. The spatial discretization is based on Bezier finite elements of arbitrary order combined with a toroidal Fourier expansion. The use of fully implicit time evolution methods avoids overly restrictive time--stepping conditions. The code uses a hybrid MPI-OpenMP parallelization.

\begin{figure}[htbp!]
    \begin{subfigure}{0.75\linewidth}
        \centering
        \includegraphics[width=0.9\linewidth]{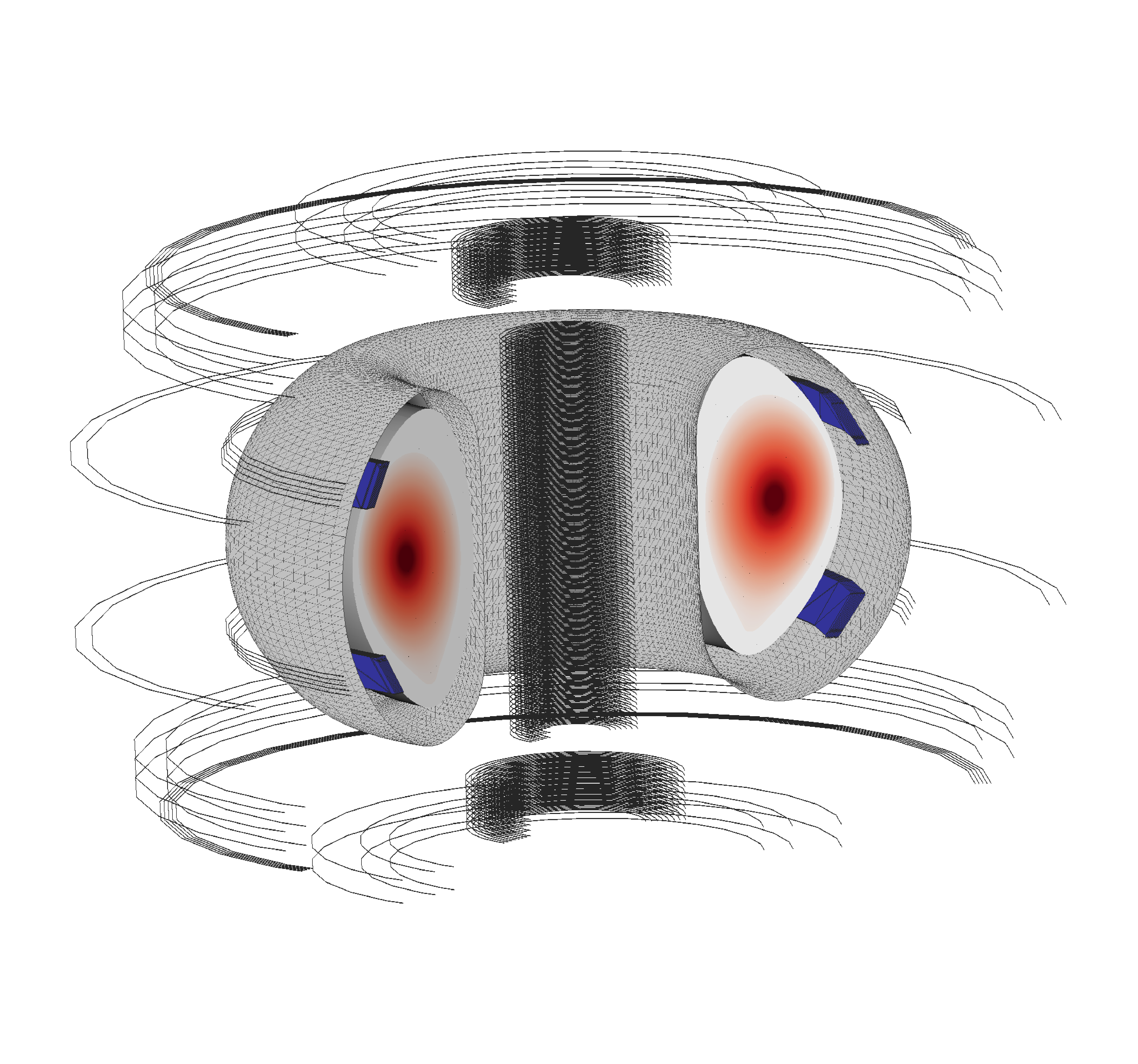}
        \caption{\label{fig:sfig01a}}
    \end{subfigure}%
    
    \begin{subfigure}{0.75\linewidth}
        \centering
        \includegraphics[width=0.9\linewidth]{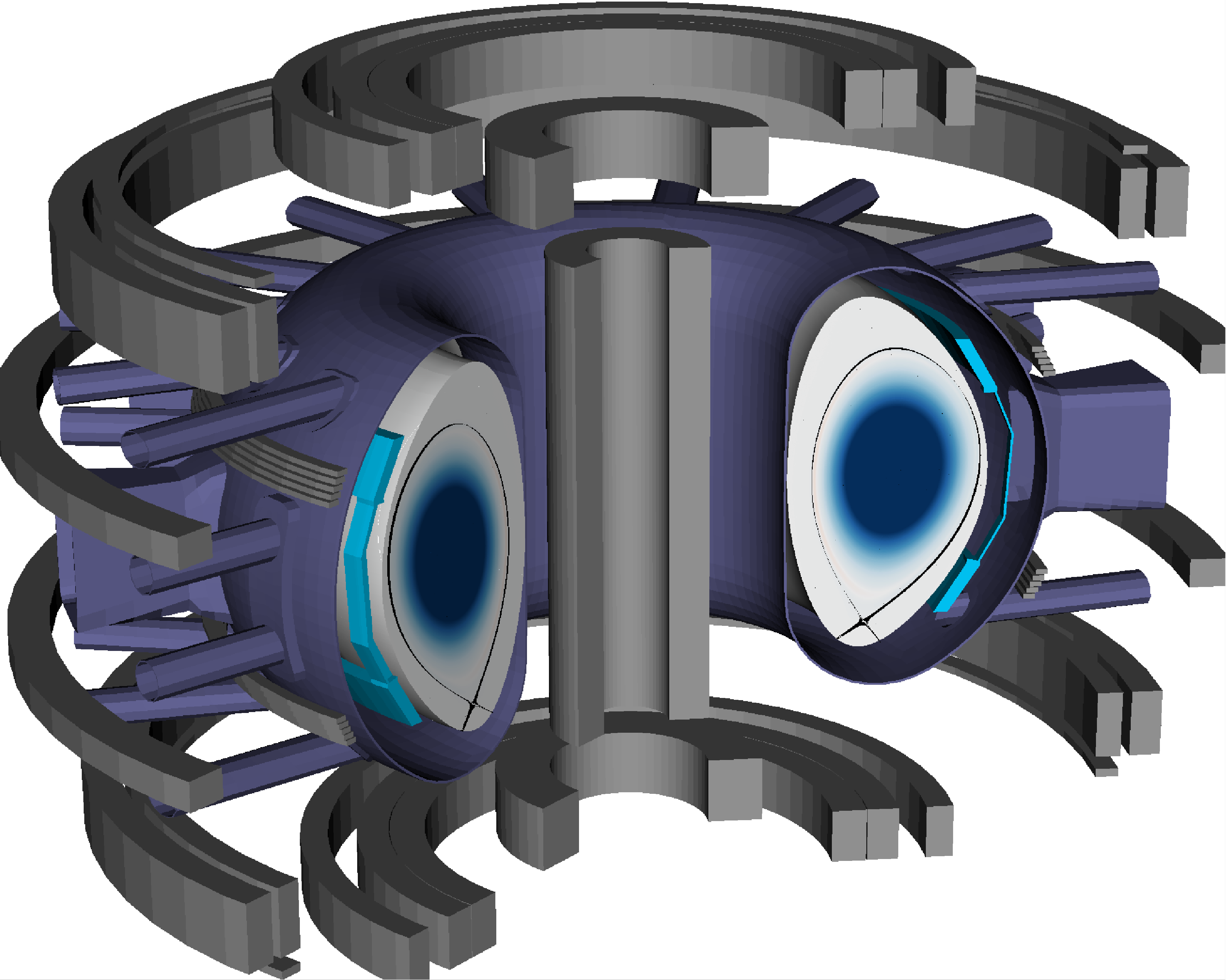}
        \caption{\label{fig:sfig01b}}
    \end{subfigure}%
    \caption{\footnotesize Examples of geometrical models for the ASDEX Upgrade (AUG) Tokamak, adopted in the couplings of JOREK with the two resistive wall codes considered here: panel~\ref{fig:sfig01a} shows the 3D thin resistive wall approximation from STARWALL; panel~\ref{fig:sfig01b} shows the 3D volumetric modeling from CARIDDI}
\label{fig:sfig01}
\end{figure}

Running JOREK standalone implies using an ideal wall boundary condition, considering all magnetic perturbations vanish at the boundary. Nevertheless, finite perturbations across the plasma boundary and the interaction of the plasma dynamics with conducting structures can be highly relevant for capturing the processes of interest. To achieve this, JOREK can leverage the calculations of external codes, STARWALL or CARIDDI~\cite{holzl2012coupling,isernia2023self}, which describe non-axisymmetric conducting structures and their interaction with the plasma. Those structures, including the walls and the coils, are discretized using a thin wall approximation in STARWALL and a volumetric description in CARIDDI, see Figure~\ref{fig:sfig01} for examples. JOREK does not discretize the region outside its domain but, instead, uses a Green's functions approach. Therefore, large dense matrices are calculated in the resistive wall codes, which allow JOREK to evolve wall and coil currents in time while calculating the magnetic field at the boundary of the computational domain self-consistently. Inside JOREK, those matrices and the boundary condition calculation are distributed across MPI tasks over the available CPUs~\cite{mochalskyy2016parallelization}. However, simulations with highly resolved wall models or high resolutions inside the plasma domain can be limited by memory consumption and computational costs.

The goal of the developments reported in this article is to explore the possibility of compressing the response matrices adopted in the JOREK free--boundary and resistive wall extension. As a first step in this task, a simple compression method has been chosen and tested against different scenarios of MHD instabilities, to understand its capabilities and limitations.

The rest of the manuscript is organized as follows. In section~\ref{sec:Impl}, details of the implementation for the matrix compression will be given, describing the chosen techniques, the terminology adopted for expressing compression and memory savings, and the present status of the work. Then, in section~\ref{sec:Sim}, the results of the simulations validate the correct implementation of the methods and prove the first demonstrations for the application of the matrix compression techniques to two different relevant kinds of plasma instabilities. Finally, in section~\ref{sec:Conc}, the implementation and verification are summarized and an outlook to future work is given.

\section{Implementation}
\label{sec:Impl}

\subsection{Matrix Compression}
\label{ssec:MatComp}

In the Linear Algebra literature, there are methods for obtaining matrix compression and controlling the accuracy of the result, with different assumptions on the structure and the features of the matrix itself. One of the less restrictive methods is based on the Singular Value Decomposition (SVD) (see e.g. ref.~\cite{trefethen2022numerical}). This factorization technique allows one to write a given matrix, $A$, with $m$ rows and $n$ columns, as

\begin{equation}
\label{eqsvd}
A = U \cdot \Sigma \cdot V^{T},
\end{equation}

being $U$ and $V^{T}$ the left and right orthogonal matrices, respectively, and $\Sigma$ the diagonal matrix of the singular values. Although the representation of (\ref{eqsvd}) is not unique, one always exists with the diagonal elements of $\Sigma$ in decreasing order. Denoting by $k_{A}$ the rank of $\Sigma$, i.e. the number of non--zero singular values, it is then clear that $U$ would have $m$ rows and $k_{A}$ columns, while $V^{T}$ would have $k_{A}$ rows and $n$ columns, and that

\begin{equation}
\label{relk}
k_{A} \leq \min(m,n).
\end{equation}

Then, instead of storing the matrix $A$ in memory, one could store the two matrices given by $(U \cdot \Sigma)$ and $V^{T}$, corresponding to a total of $(n+m) \cdot k_{A}$ numbers instead of $m \cdot n$. Selecting one value $k < k_{A}$, thus neglecting the $(k_{A} - k)$ smallest singular values, leads to compressing the required memory for storing the matrix $A$ when

\begin{equation}
\label{relmemsav}
(m+n) k < m n.
\end{equation}

In other words, indicating with $r_r$ the chosen rate of retained singular values, such that

\begin{equation}
\label{rr}
k = r_r \min(m,n),
\end{equation}

using (\ref{rr}) in (\ref{relmemsav}), memory is saved provided

\begin{equation}
\label{save}
r_r < \frac{\max(m,n)}{m+n}.
\end{equation}

The compressed SVD representation of matrix $A$ with a rank $k < k_A$ is often called the truncated SVD.

Additionally, it is useful to define a rate of memory gain, $r_{mg}$, of a given compressed matrix to measure the efficiency of compression in applications. Computing the percentual relative difference between the dimension of the original matrix and the compressed one, $r_{mg}$ reads

\begin{equation}
\label{mgr}
r_{mg} = \left( 1.0 - \frac{ (m+n) r_r}{\max(m,n)} \right) * 100 \%.
\end{equation}

Of course, $r_{mg}$ can be negative, which is expected when the condition (\ref{save}) is not verified. For example, when $r_r = 1.0$, the matrix is not compressed but only factorized, therefore $r_{mg} < 0$. 

In terms of accuracy, finally, the Eckart--Young theorem proves that the truncated SVD is an optimal approximation of a given matrix at fixed rank when computing the error via the Frobenius norm (see~\cite{stewart1993early} for a review and~\cite{eckart1936approximation} for the original reference).

\subsection{Matrices chosen for the implementation}
\label{ssec:MatChosen}
The coupling of JOREK to STARWALL or CARIDDI introduced in section~\ref{sec:Intro} adopts the virtual casing principle to express the interaction between the plasma and eddy currents in the wall by a set of static matrices. The reader is referred to Appendix C of \cite{such2018free} for a general description of all the geometrical matrices involved in the calculations.

In the present study, the dense matrices $M^{ey}$ and $M^{ye}$ of ref.~\cite{holzl2012coupling, isernia2023self} were chosen, which describe the mutual interactions between the wall and the plasma contained in it. Without entering into formal definitions, it is worth recalling the physical meaning of those matrices. On the one hand, the matrix $M^{ye}$ relates the magnetic flux variations at the JOREK boundary to current variations in the external conductors. On the other, the matrix $M^{ey}$ provides information on the tangent magnetic field to the JOREK boundary produced by the external currents when a superconducting shell at the JOREK boundary is considered.

Denoting by $n_{w}$ the degrees of freedom (DoF) of the wall and by $n_{bd}$ the DoF at the boundary of the plasma region, $M^{ey}$ is of dimension ($n_{bd} \times n_{w}$), while $M^{ye}$ is of dimension ($n_w \times n_{bd}$) (see ref.~\cite{holzl2012coupling} for the details).
As can be seen, both matrices have $n_w \cdot n_{bd}$ entries. In many cases, these matrices dominate the memory consumption when very detailed wall models are used in the simulations. Following the geometrical finite elements representation adopted in JOREK and recalled in section~\ref{sec:Intro}, one can write

\begin{equation}
\label{dofbound}
n_{bd} = n_{B} n_{F} = n_{B} ( 1 + 2 k_{F}),
\end{equation}

being $n_{B}$ the DoF of the poloidal Bezier elements, $n_{F}$ the toroidal Fourier harmonics, where sine and cosine components are counted separately (as in the JOREK code), and $k_{F}$ the Fourier mode number.
Therefore, one immediate way of changing the dimension $n_{bd}$ of the chosen matrices, would be to increase the number of toroidal harmonics to be considered, as reported in section~\ref{ssec:MatComp}, even without varying $n_{w}$. This gives a fast way to test the implementation with different sizes of the matrices.

To clarify the importance of compressing the selected matrices, it is sufficient to look at the state--of--the--art of HPC hardware, numerical modeling of the free--boundary and resistive wall extension, and concrete dimensions of target Tokamaks. At first, looking at the HPC systems within the Top500 list (as of June 2024 - refer to~\cite{top500website} for the list, although the RAM per node value info can be found separately in each system's documentation), the available RAM on one node may span from $1e2$ up to $1e3$ $GB$; one could take $512$ $GB$ as a reference for this quantity, which can be filled with a totality of $64e9$ numbers in double precision. From the numerical modeling and target Tokamaks points of view, in the case of ITER or DEMO, greatly overcoming the dimensions of any existing device, typical values for the DoF of the passive structures on the walls, $n_w$, range from $1e5$ to several times $1e6$. In addition, well--resolved plasma boundaries require values of $n_{bd}$ ranging from $1e3$ to $1e5$ DoF (the bigger values are in the case of highly--resolved poloidal regions). Finally, the response matrices are dense, as already mentioned in the introduction and the present section, therefore they cannot be represented in any memory--efficient formats. Consequently, the two response matrices considered, which count $n_w \times n_{bd}$ entries, may each require up to $1e11$ numbers for the target Tokamaks to be modeled. From all these considerations, it follows that data compression is the only possible way to treat arbitrarily accurate full 3D geometrical models of ITER or DEMO in the free--boundary and resistive wall regime.

\subsection{Code Development and Optimization}
\label{ssec:Opt}
The SVD is performed with the Scalable Linear Algebra PACKage (ScaLAPACK, see ref.s~\cite{blackford1997scalapack, scalapackWebsite}), through the subroutine \texttt{pdgesvd}. As the matrices are static, the compression is only required once before the start of the JOREK simulation. The compression is performed by the newly developed \texttt{compress\_response} program. In detail, the \texttt{compress\_response} program should be executed after the production of response matrices from STARWALL or CARIDDI and before the simulation to be performed with JOREK. A simple sketch of this implementation is presented in algorithm~\ref{alg:cr}. Here, after the singular values are determined by \texttt{pdgesvd}, the smallest values are eliminated based on the $r_r$ of the user, and the truncated SVD is thus obtained.

\begin{algorithm}
\caption{\texttt{compress\_response}}\label{alg:cr}
\begin{algorithmic}
\State \textbf{Input:} response file, selection of matrices to be compressed, value $r_r$ for each matrix to be compressed
\State $A$ $\gets$ read response file
\If{$A$ to be compressed}
    \State $\left( U, \Sigma, V^T \right) \gets$ call \texttt{pdgesvd} with $A$ \Comment{SVD}
    \State $n_{\Sigma} \gets$ \texttt{size}($\Sigma$)
    \State $n_{\Sigma}^{compr} \gets \left( n_{\Sigma} \cdot r_r \right)$
    \State $\Sigma^{compr} \gets \Sigma \left( 1 \dots n_{\Sigma}^{compr} \right)$ \Comment{compression}
    \State keep $\left( \left( U \cdot \Sigma^{compr} \right), V^T \right)$
\ElsIf{$A$ not to be compressed}
    \State $A$ unchanged
\EndIf
\State write output
\State \textbf{Output:} new response file
\end{algorithmic}
\end{algorithm}

For debugging purposes, it is also possible to print out re--combined versions of the factorized matrices, to print out the SVD analysis details, and to compress only the non--axisymmetric part of the matrix.

The isolation of the compression in a separate program allowed limiting the number of modifications to the original JOREK code. Nevertheless, the JOREK code had to be adapted to work with the chosen matrices in a factorized format. In references~\cite{holzl2012coupling,isernia2023self,such2018free}, it can be seen that both $M^{ey}$ and $M^{ye}$ of section~\ref{ssec:MatChosen} are used in algebraic computations involving matrix--vector products. Those operations were adopted to calculate the derived response matrices temporarily stored in the memory. Eliminating such derived response matrices and implementing a new generalized matrix--vector product instead save additional Random Access Memory during the code execution.

It is relevant to estimate how much the computational cost is modified when considering factorized matrices. In particular, the matrices $M^{ey}$ and $M^{ye}$ are used to update the wall currents and to compute the vacuum boundary integral for obtaining the magnetic field terms (see Section 3.2 of~\cite{holzl2012coupling}). There are $5$ to $7$ matrix--vector products per time iteration, depending on whether the previously used time--step is kept or changed. Therefore, estimating the computational cost of such a single calculation can give a general idea of the overall number of operations. Here, in general, one finds that the number of scalar multiplications is

\begin{equation}
\label{cost}
cost_{aggr} = m_A \cdot n_A,    
\end{equation}

when the matrix $A$ is aggregated. However, when considering the factorization of (\ref{eqsvd}) with a truncated SVD retaining $k$ singular values as in (\ref{rr}), then the overall cost is

\begin{equation}
\label{costcompr}
cost_{compr} = \left( m_A \cdot k \right) + \left( k \cdot n_A \right).
\end{equation}

The direct comparison of the two costs provides the condition on the number of retained singular values to have computational cost savings:

\begin{equation}
\label{savecost}
k < \frac{m_A \cdot n_A}{\left( m_A + n_A \right)}.
\end{equation}

Given that those operations happen at each time iteration, it was crucial to be sure that the factorization of the matrix does not deteriorate the overall performance. To this end, hybrid \texttt{MPI}/\texttt{OpenMP} parallelization was exploited, factorized matrices were distributed across MPI ranks, and the nested loops involving them have been adapted. Those optimizations allowed retaining the same computational time when adopting factorized but not compressed matrices or their original aggregated version.

\begin{figure}[h!]
    \centering
    \includegraphics[width=0.5\linewidth]{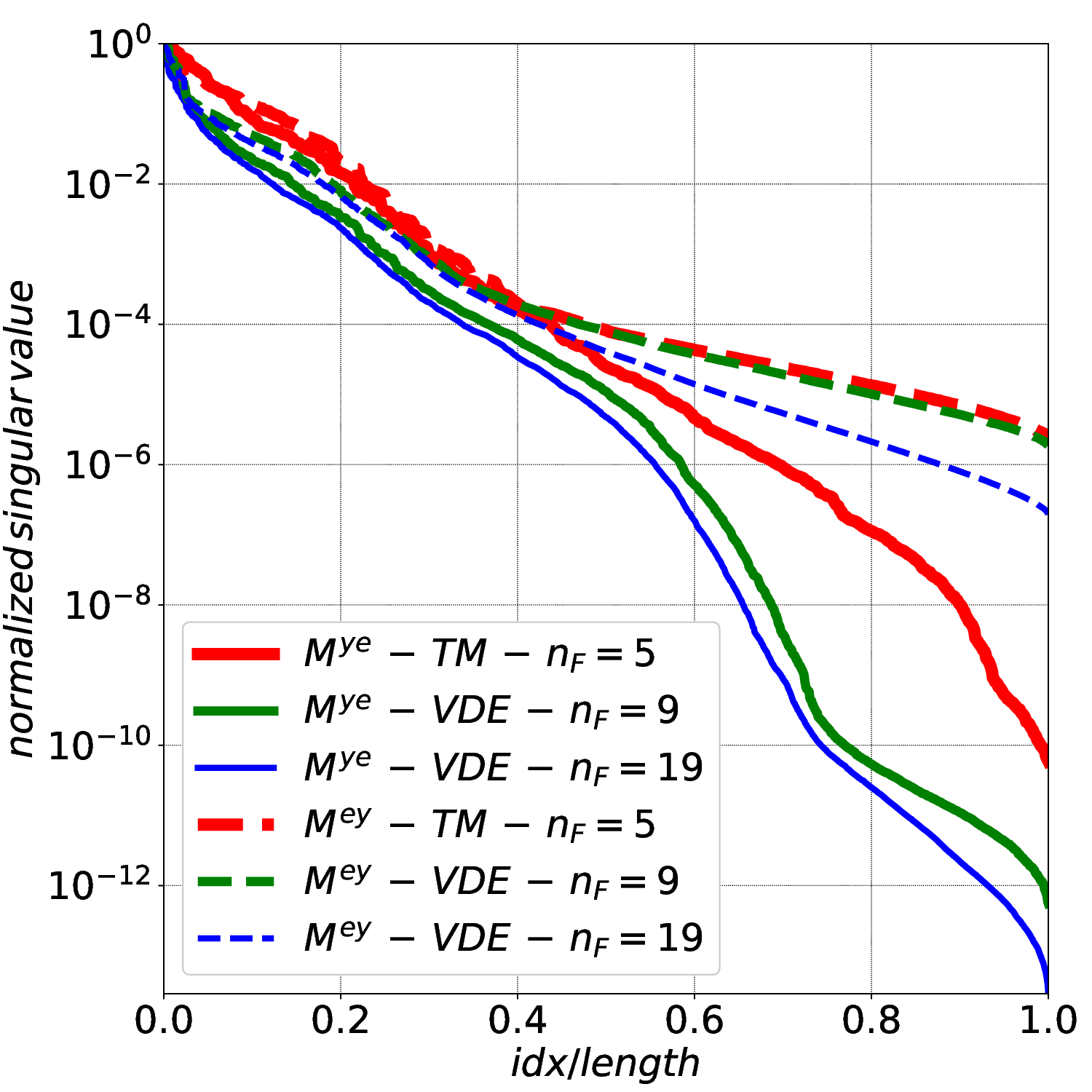}
    \caption{\footnotesize\label{fig:fig02} Spectral analysis of the matrices for the Tearing Mode (TM) and Vertical Displacement Event (VDE) test cases, respectively presented in sections~\ref{ssec:TM} and~\ref{ssec:VDE}. The line styles and colors (in the online version) are explained in the legend}
\end{figure}

To conclude the present section, an insight into the practical meaning of equations (\ref{cost}), (\ref{costcompr}), and (\ref{savecost}) is provided by the spectral analysis reported in figure~\ref{fig:fig02}. Given that the normalized singular values of matrix $M^{ye}$ are slightly smaller than the ones of $M^{ey}$, in the adopted discretized geometrical model, the former  is expected to show better compressibility than the latter.

\section{Verification and Testing}
\label{sec:Sim}

The JOREK code can simulate various kinds of plasma instability that may occur in a magnetically confined plasma~\cite{hoelzl2021jorek,hoelzl2024jorek}. To test the implementation presented in section~\ref{sec:Impl}, two such scenarios were selected, namely the Tearing Mode (TM) instability and the Vertical Displacement Event (VDE). The simulation setups and results of compression tests are shown in the following sections. Note that we used comparably small test cases here. In the future, computationally more expensive production simulations will be done after the modifications have been merged into the main code version. Regarding the test methodology, the compression capability was estimated by adopting several values of $r_r$ and verifying if significant quantities could be computed correctly as in the uncompressed case. When trying to define a threshold for an optimal value, only one iteration of the bisection method was attempted between one successful and one failing value of $r_r$ (e.g. in the case of $r_r = 0.15$ in section~\ref{ssec:TM}). Of course, the minimum value of $r_r$ found in this way is not to be considered a global minimum but rather a proof of concept of the technique adopted.

\subsection{Tearing Mode Instability}
\label{ssec:TM}

The Tearing Mode (TM) instability is a well--known phenomenon in the field of MHD (see for example~\cite{hazeltine1978introduction} for a classical review on this topic). Despite its theoretical conception being initially driven by astrophysical argumentations, the relative scenario was recovered and deeply studied in the experimental tokamak plasma physics field. This instability modifies the magnetic topology by reconnecting magnetic field lines such that the so--called magnetic islands are formed. Under certain circumstances, TMs can grow to significantly large amplitude and cause major disruptions, during which the confinement of the plasma is lost in an abrupt way causing large heat and electromagnetic loads for the device.

\begin{table}
\caption{\footnotesize\label{tab:TM}Details of TM instabilities simulations performed for this work. The left column indicates the rate of retained singular value adopted; here a ``$-$'' character refers to the case of the original aggregated matrix. The central column reports $s$, the resulting size in $GiB$ of the matrix $M^{ey}$ or $M^{ye}$. The right column reports the rate of memory gain defined in~(\ref{mgr}).\normalsize}
\begin{center}
\begin{tabular}{@{}ccc}
\br
$r_r$&$s \left[ GiB \right]$&$r_{mg} \left[ \% \right]$\\
\mr
$-    $&$0.121 $&$0.0   $\\
$1.0  $&$0.126 $&$-4.13 $\\
$0.75 $&$0.0944 $&$21.98$\\
$0.5  $&$0.0629 $&$48.02$\\
$0.25 $&$0.0315 $&$73.9 $\\
$0.2  $&$0.0252 $&$79.17$\\
$0.15 $&$0.0189 $&$84.38$\\
$0.1  $&$0.0126 $&$89.59$\\
$0.05 $&$0.00629$&$94.80$\\
$0.025$&$0.00315$&$97.40$\\
\br
\end{tabular}\\
\end{center}
\end{table}

\begin{figure}[h!]
    \centering
    \includegraphics[width=0.5\linewidth]{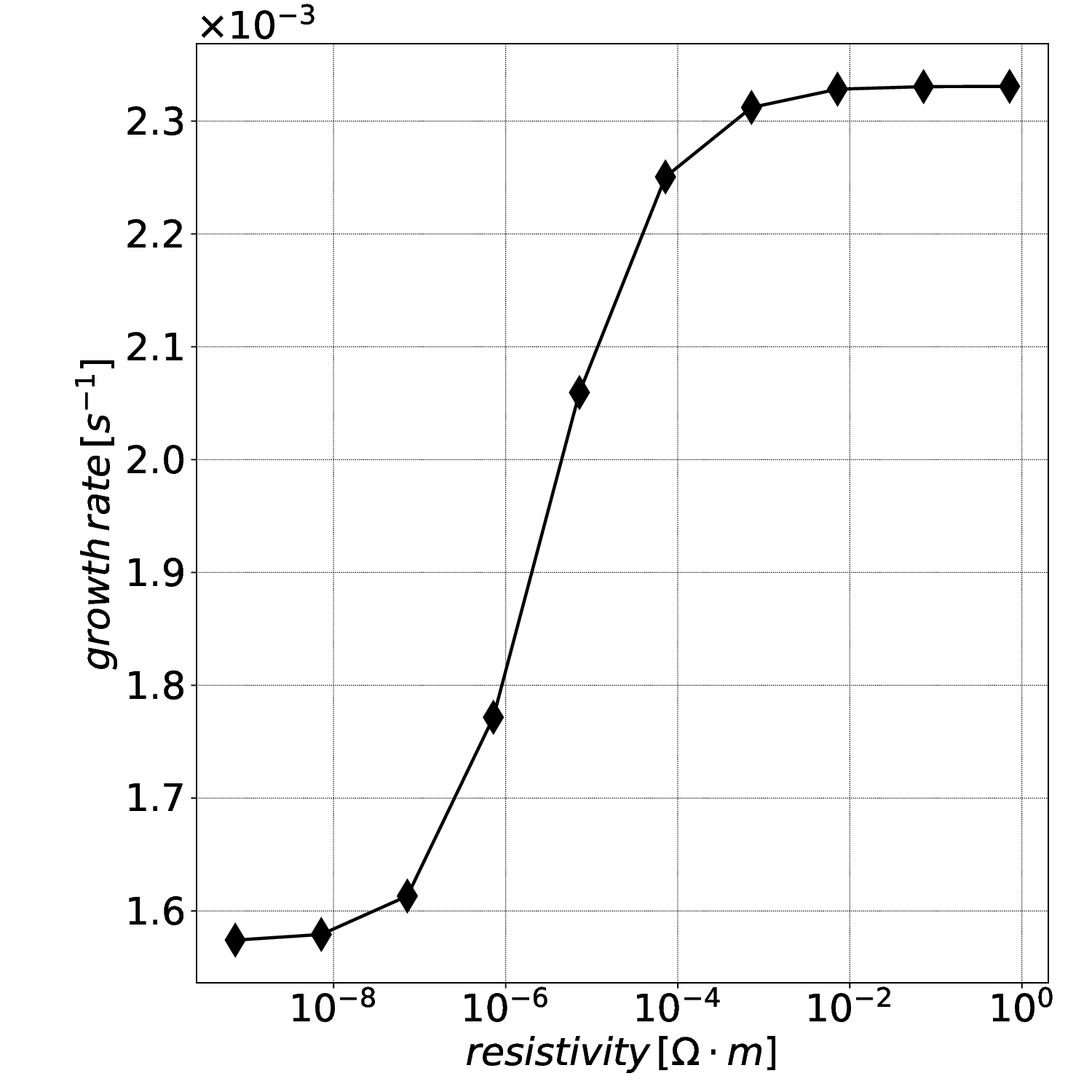}
    \caption{\footnotesize\label{fig:fig03}Dependence of the linear growth rate of the TM instability on the wall resistivity for the test case of section \ref{ssec:TM}. The scan covers the whole range from the ideal wall limit to the no-wall limit}
\end{figure}

The test case considered here is linearly unstable to a tearing mode with dominant toroidal mode number $1$ and the vacuum response matrices are calculated by the CARIDDI code. The geometrical structure adopted is similar to what is shown in figure~\ref{fig:sfig01b}. The axisymmetric component of the magnetic field (with mode number $0$) is not included in the response matrices, such that an ideal wall boundary condition is applied for the axisymmetric component. This avoids dealing with vertical plasma instabilities, which are addressed separately in the following section.  The variation of the linear growth rate of the TM as a function of the wall resistivity is reported in figure~\ref{fig:fig03}. The nominal wall resistivity used in further studies below has a value of $0.72 \times 10^{-6} \, \left[ \Omega \cdot m \right]$. The $n=0$ magnetic flux stays approximately constant during the evolution: this was observed by checking that the absolute relative difference with respect to its maxima between the saturated and equilibrium instants is below $4e-4$.

\begin{figure}[h!]
    \begin{subfigure}{.5\textwidth}
        \centering
        \includegraphics[width=0.9\linewidth]{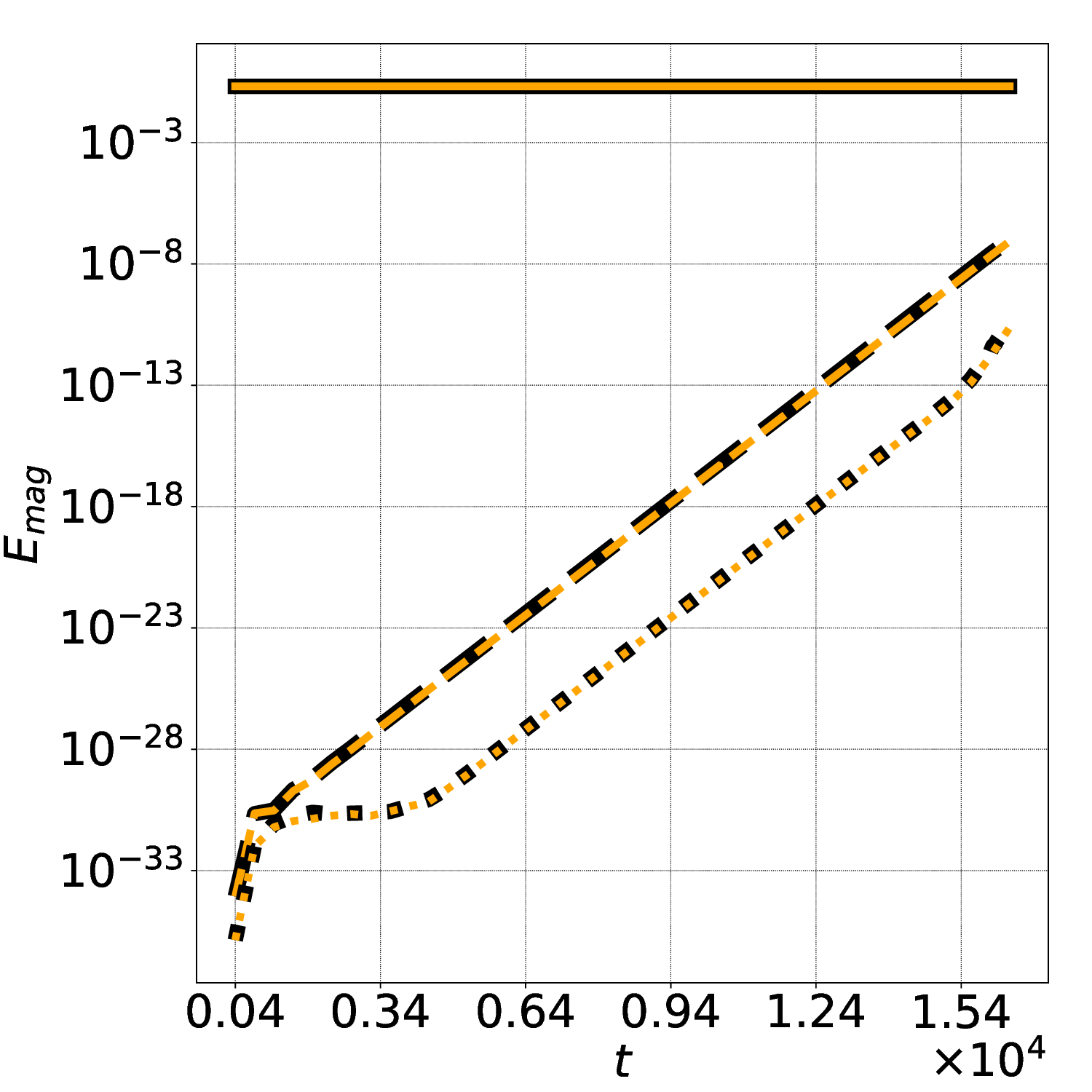}
        \caption{\label{fig:sfig04a}}
    \end{subfigure}%
    \begin{subfigure}{.5\textwidth}
        \centering
        \includegraphics[width=0.9\linewidth]{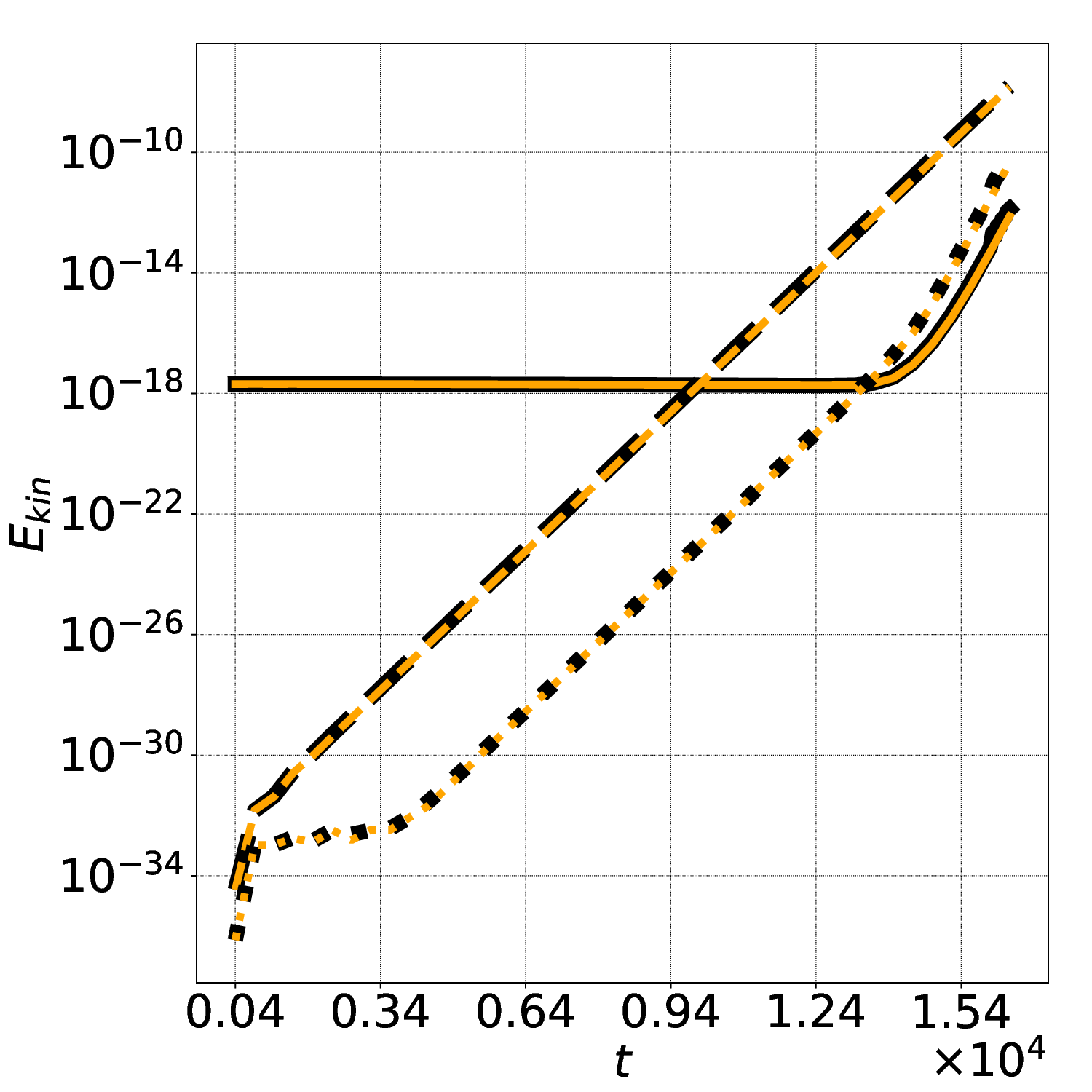}
        \caption{\label{fig:sfig04b}}
    \end{subfigure}
    \caption{\footnotesize\label{fig:fig04}Comparison of the time evolution of the Fourier harmonics of magnetic (panel~\ref{fig:sfig04a}) and kinetic (panel~\ref{fig:sfig04b}) energy in the simulations of the TM instability. Solid curves stand for $k_F=0$ harmonic, dashed for $k_F=1$, and dotted for $k_F=2$. Black curves refer to the results from the original aggregated version of the matrices. Orange curves refer to the results from the factorized but uncompressed matrices ($r_r = 1.0$)}
\end{figure}

Before looking into the compression efficiency, the correctness of the implementation is tested. For this purpose, a case with $n_w = 20322$, $n_B = 160$, $n_F = 5$ is simulated both with the uncompressed matrices and with the factorized matrices without compression ($r_r=1$). The comparison of the magnetic and kinetic energies in the different toroidal harmonics shown in Figure~\ref{fig:fig04} shows perfect agreement.

\begin{figure}[h!]
    \begin{subfigure}{.5\textwidth}
        \centering
        \includegraphics[width=0.9\linewidth]{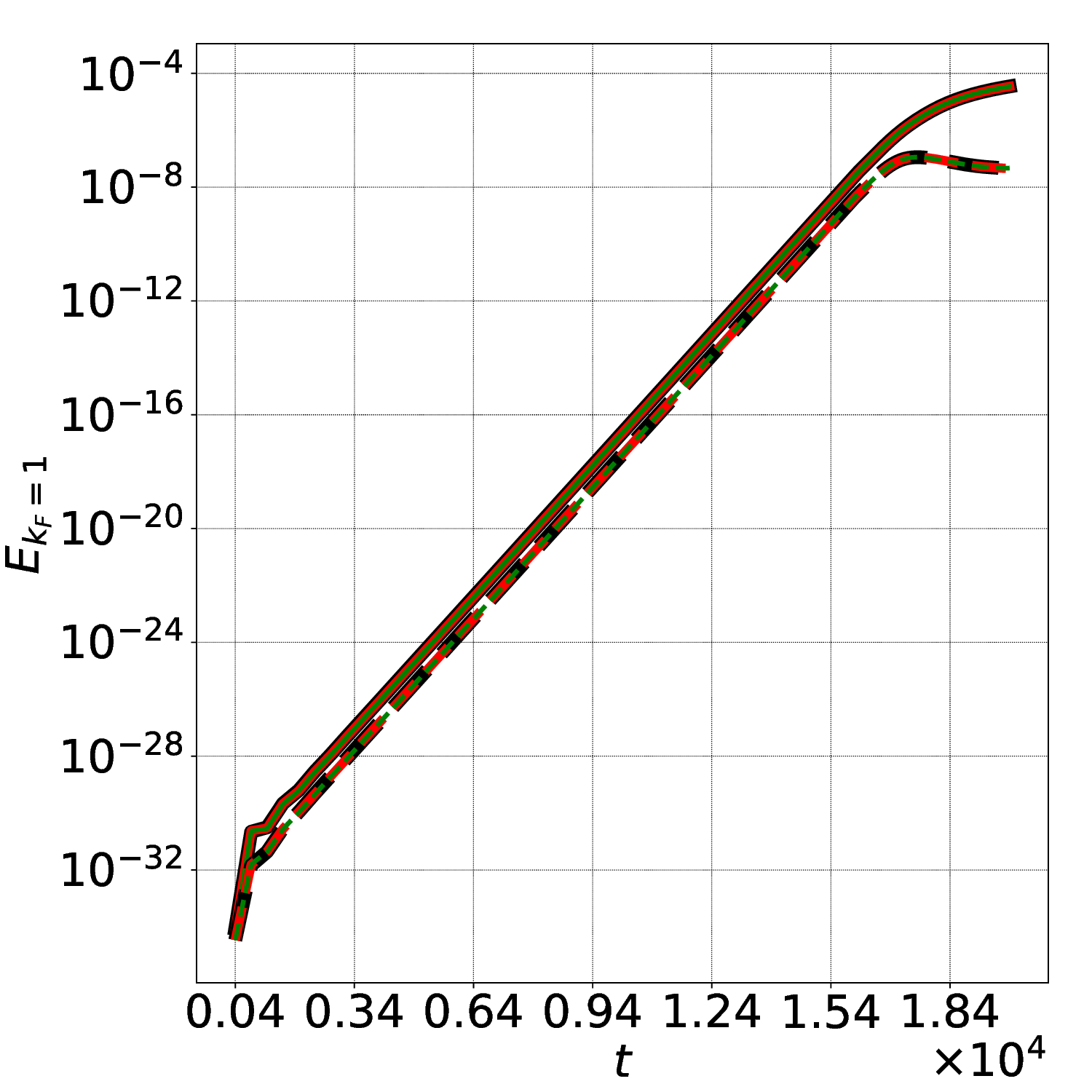}
        \caption{\label{fig:sfig05a}}
    \end{subfigure}%
    \begin{subfigure}{.5\textwidth}
        \centering
        \includegraphics[width=0.9\linewidth]{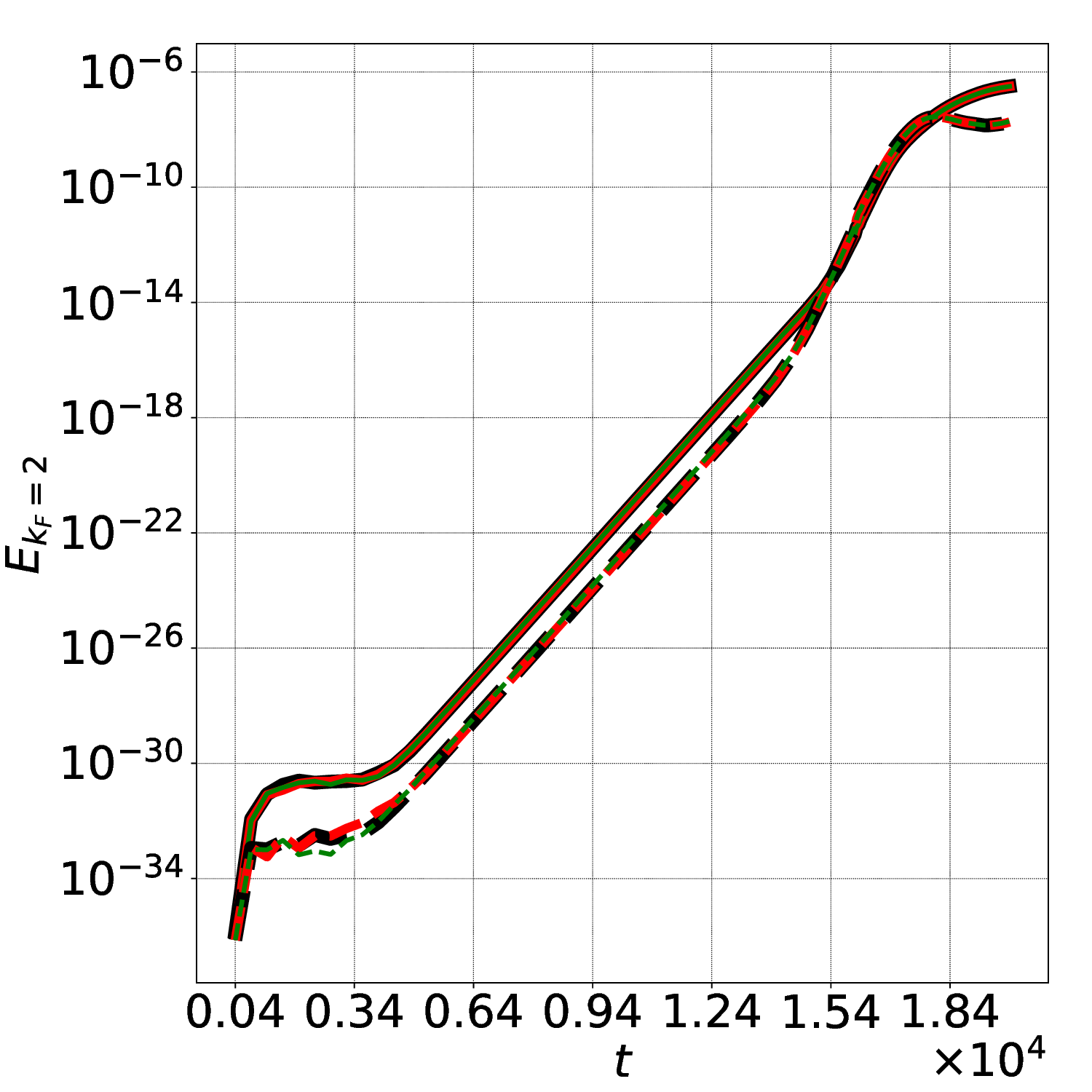}
        \caption{\label{fig:sfig05b}}
    \end{subfigure}
    \caption{\footnotesize\label{fig:fig05}Comparison of the time evolution of the $k_F = 1$ (panel~\ref{fig:sfig05a}) and $k_F = 2$ (panel~\ref{fig:sfig05b}) mode numbers of the magnetic and kinetic energy, for the TM instability simulations, up to the non--linear saturation. The magnetic energy is depicted with solid curves and kinetic energy with dashed curves. The black color is used for $r_r = 1.00$, red for $r_r = 0.20$, and green for $r_r = 0.15$}
\end{figure}

The efficiency of the compression is now tested using different retention rates in the factorized matrices. In table~\ref{tab:TM}, the values adopted for $r_r$ are reported together with the size $s$ of the corresponding truncated SVD matrices and the relative memory usage reduction $r_{mg}$ given by~(\ref{mgr}), computed based on the original size. It is worth noting that the first line corresponds to the same simulation performed with aggregated and uncompressed matrices and should give matching results with $r_r = 1.0$ for validation.
In the tests performed, accurate results could be obtained with retention rates down to $r_r = 0.15$ as shown in Figure~\ref{fig:fig05}. With smaller values of $r_r$ the results were instead inaccurate. 

\begin{figure}[h!]
    \centering
    \includegraphics[width=0.5\linewidth]{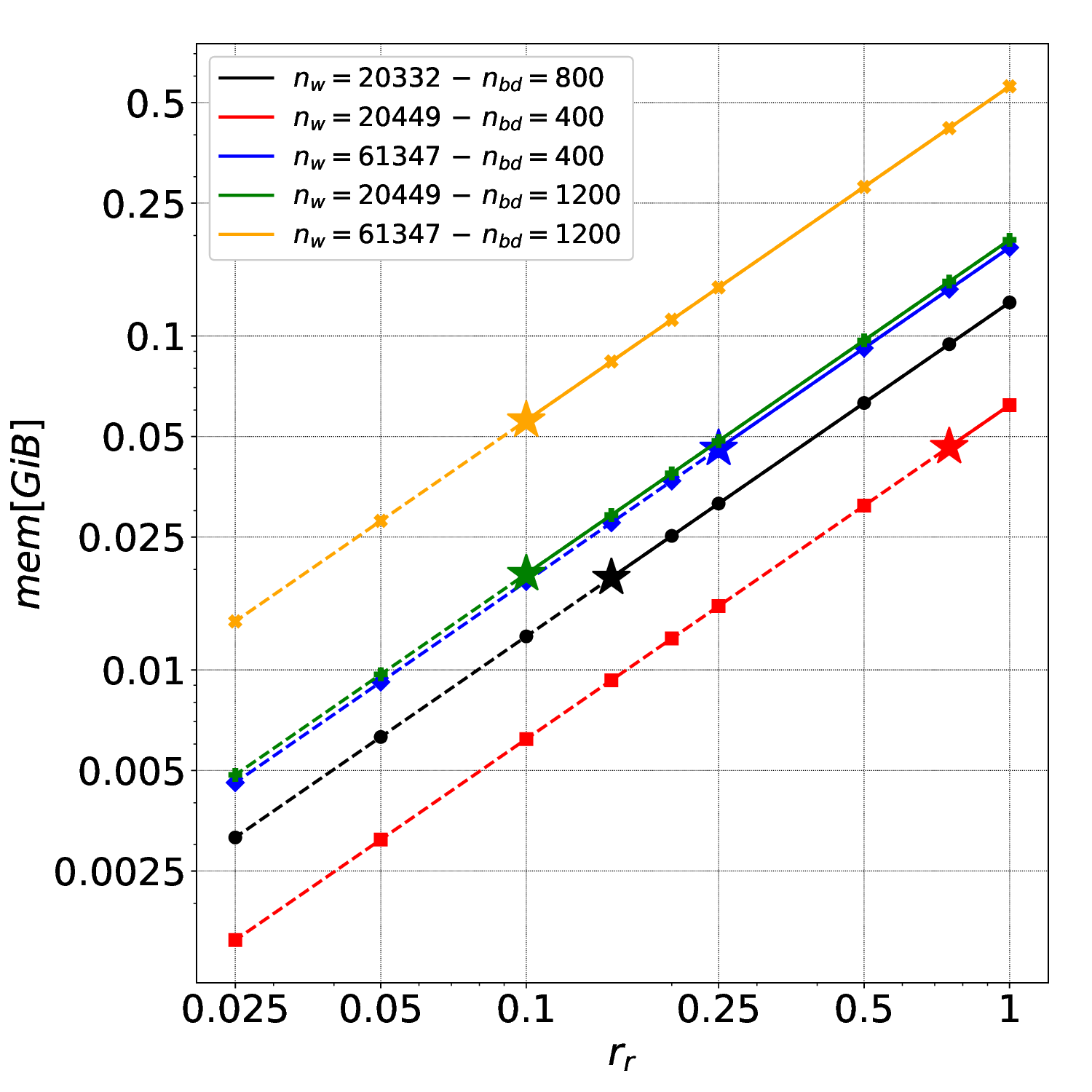}
    \caption{\footnotesize\label{fig:fig06}Memory occupation of matrices in $\left[ GiB \right]$ against the rate of retained singular values for TM instability simulations with different wall ($n_w$) and plasma ($n_{bd}$) resolutions, plotted in double logarithmic scale. On each line, the following convention is used: the solid style indicates a region where accurate results were obtained; the dashed style indicates inaccurate results; markers locate tested values of $r_r$; a filled star locates the minimum value of $r_r$ for which accurate results were obtained, therefore the maximum compression reachable. The black curve corresponds to the resolution of the first set of tests relative to table~\ref{tab:TM}, where the response file was produced using the CARIDDI code. Other colors (in the online version) correspond to results from response files produced with STARWALL, varying the number of boundary elements as indicated in the legend.}
\end{figure}

As a general statement, it is worth noting that the efficiency of the compression of a matrix is expected to grow with increasing dimensions of the matrix itself. Therefore, a series of tests with varying resolutions of the boundaries on the wall ($n_w$) and the poloidal section of the plasma ($n_{B}$) has been performed, keeping fixed $n_F=5$. These scans are performed with the STARWALL instead of the CARIDDI code for simplicity, thus adopting a geometrical structure like what is shown in figure~\ref{fig:sfig01a}, starting from a base resolution of $n_w=20449$, $n_{bd}=400$ and increasing either the resolution of the wall or the plasma independently. The results shown in figure~\ref{fig:fig06} indeed show better compressibility for larger problem sizes. However, it is also worth noting that the maximum compressibility obtainable via truncated SVD is limited by the minimum between the number of rows and columns of the matrix, as understandable from equations (\ref{eqsvd}) and (\ref{relk}). In the studied cases, this limit is always given by the number of boundary elements on the plasma side.

\subsection{Vertical Displacement Event (VDE)}
\label{ssec:VDE}

VDE refers to a loss of control of the vertical positioning of the magnetically confined plasma inside a tokamak (see~\cite{clauser2019vertical} and reference therein for a recent report of simulations of this instability with the ITER geometry). The following loss of thermal and magnetic energy can reduce the machine's lifetime, therefore it is crucial to understand its dynamics with the help of numerical simulations. 

In references~\cite{schwarz2020vertical, schwarz2023experiments}, 
it is possible to see how such a phenomenon can be effectively set up with the STARWALL code and the AUG geometry, to be simulated with the JOREK code, taking into account initial perturbation on the coil current. Moreover, reference~\cite{artola20213d} reports VDE simulations to benchmark comparison between the codes M3D-C$^1$, NIMROD, and JOREK, with a simplified geometry. In addition, reference~\cite{isernia2023self} contains results of simulations of VDE where the response was computed adopting CARIDDI. 
The VDE simulations of the present section closely follow the setup adopted in already published work, described in \cite{schwarz2020vertical}, \cite{isernia2023self}. It consists of the following steps:

\begin{enumerate}
    \item[1.] a 2D axisymmetric phase, taking into account only $n_F=0$ harmonic until the magnetic axis reaches the vertical position Z = -0.1 m;
    \item[2.] a full 3D restart taking into account all the chosen $n_F$ harmonics;    
    \item[3.] restart at the Thermal Quench (TQ) phase with constant perpendicular diffusivities, no particle source, and decreasing the
    time--step;
    those settings simplified the simulations, allowing their continuation, and alleviating overly restrictive limitations on the time--steps during TQ.
\end{enumerate}

\begin{table}
\caption{\footnotesize\label{tab:VDE}Details of VDE simulations performed for this work, regarding the memory required by matrix $M^{ey}$ or $M^{ye}$, indicated with $s$, and the rate of memory saving, $r_{mg}$, for each choice of $n_F$ and each value of $r_r$ chosen for compression\normalsize}
\begin{center}
\begin{tabular}{@{}ccccccc}
\br
$r_r$&$s_{n_F=1} \left[ GiB \right]$&$r_{mg}^{n_F=1} \left[ \% \right]$&$s_{n_F=9} \left[ GiB \right]$&$r_{mg}^{n_F=9} \left[ \% \right]$&$s_{n_F=19} \left[ GiB \right]$&$r_{mg}^{n_F=19} \left[ \% \right]$\\
\mr
$-   $&$0.0273 $&$0.0  $&$0.245 $&$0.0  $&$0.518$&$0.0   $\\
$1.0 $&$0.0275 $&$-7.32$&$0.265 $&$-8.1 $&$0.605$&$-16.79$\\
$0.75$&$0.0206 $&$24.54$&$0.199 $&$18.77$&$0.454$&$12.35 $\\
$0.5 $&$0.0137 $&$49.81$&$0.132 $&$46.12$&$0.302$&$41.7  $\\
$0.4 $&$0.0110 $&$59.7 $&$0.106 $&$56.73$&$0.242$&$53.28 $\\
$0.3 $&$0.00825$&$69.78$&$0.0795$&$67.55$&$0.181$&$65.06 $\\
$0.25$&$0.00687$&$74.84$&$0.0662$&$72.98$&$0.151$&$70.85 $\\
\br
\end{tabular}\\
\end{center}
\end{table}

The VDE tests of the present paper were performed by adopting the AUG geometry inside CARIDDI (refer to the geometrical structure of figure~\ref{fig:sfig01b}) with $n_w = 20322$, $n_B = 180$, and different values of $n_F$ (namely $1$, $9$, and $19$), to evaluate the effect of the compression of the chosen matrices on the accuracy of the toroidal Fourier harmonics representation. Moreover, several values of $r_r$ were initially taken into account, for both the $M^{ey}$ and $M^{ye}$ introduced in section~\ref{ssec:MatChosen}. The details of the memory sizes involved in those VDE simulations are reported in table~\ref{tab:VDE}. As expected, going from the top row to the bottom one, it is possible to see that compressing the matrices at higher rates allows for saving more memory.

\begin{figure}[h!]

    \begin{subfigure}{.5\textwidth}
        \centering
        \includegraphics[width=0.9\linewidth]{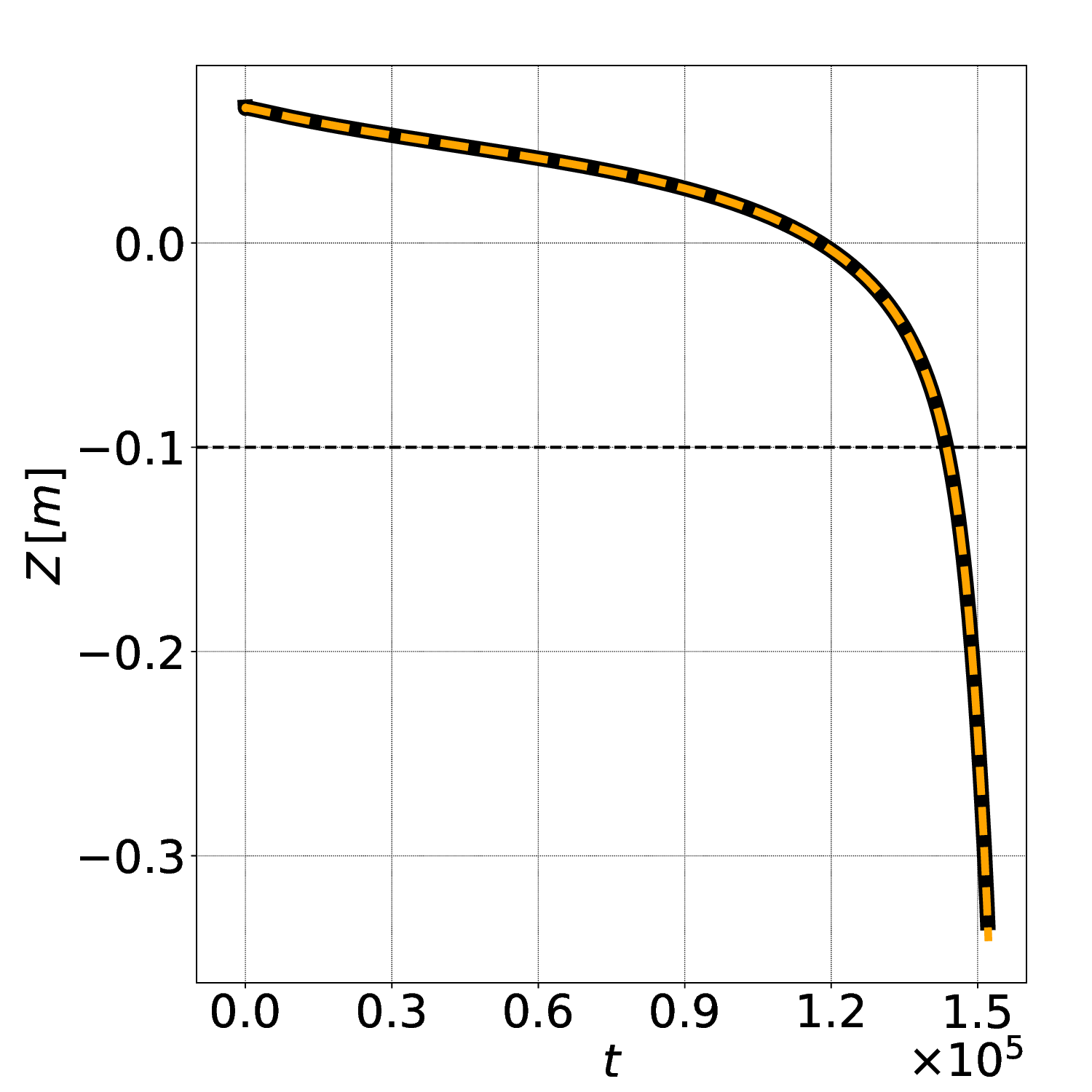}
        \caption{\label{fig:sfig07a}}
    \end{subfigure}%
    \begin{subfigure}{.5\textwidth}
        \centering
        \includegraphics[width=0.9\linewidth]{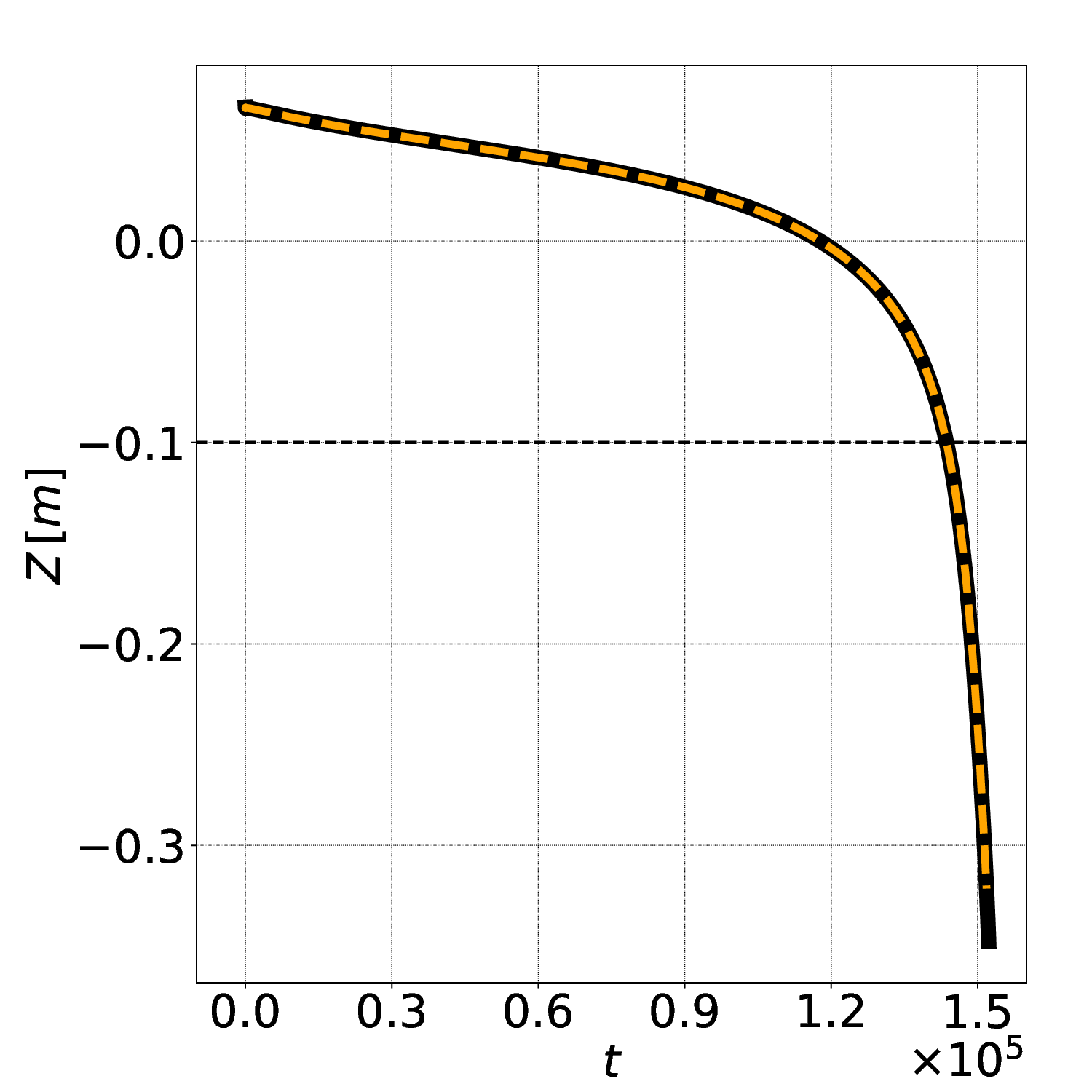}
        \caption{\label{fig:sfig07b}}
    \end{subfigure}
    \caption{\footnotesize\label{fig:fig07}Test for the verification of the implementation. Time evolution of the $Z$--coordinate of the magnetic axis. In panel~\ref{fig:sfig07a}, $n_F=9$ toroidal harmonics were used. In panel~\ref{fig:sfig07b}, $n_F=19$ toroidal harmonics were used. A dashed horizontal black line indicates the start of the 3D phase in the simulations. In the online version, the black solid curves depict the result from the original aggregated version of the matrices, while the dashed orange curves are for the factorized but uncompressed ones \normalsize}
\end{figure}

In all the simulated cases, the matrices factorized with $r_r = 1.0$ reproduced exactly the results of the original version of the matrix, thus verifying the implementation, see figure~\ref{fig:fig07}, where the time evolution of the vertical axis position is compared between the original and factorized but uncompressed matrices. 

On the other hand, all the simulations adopting $r_r < 1.0$ failed to produce accurate results, when compressing both the $M^{ey}$ and $M^{ye}$ response matrices. Although that preliminary outcome might indicate incompressibility of the matrices in the VDE test case, also the single $M^{ye}$ compression was attempted with $r_r = 0.4$, $0.6$, and $0.8$ while leaving $M^{ey}$ factorized with $r_r = 1.0$. 

\begin{figure}[h!]

    \begin{subfigure}{.5\textwidth}
        \centering
        \includegraphics[width=.9\linewidth]{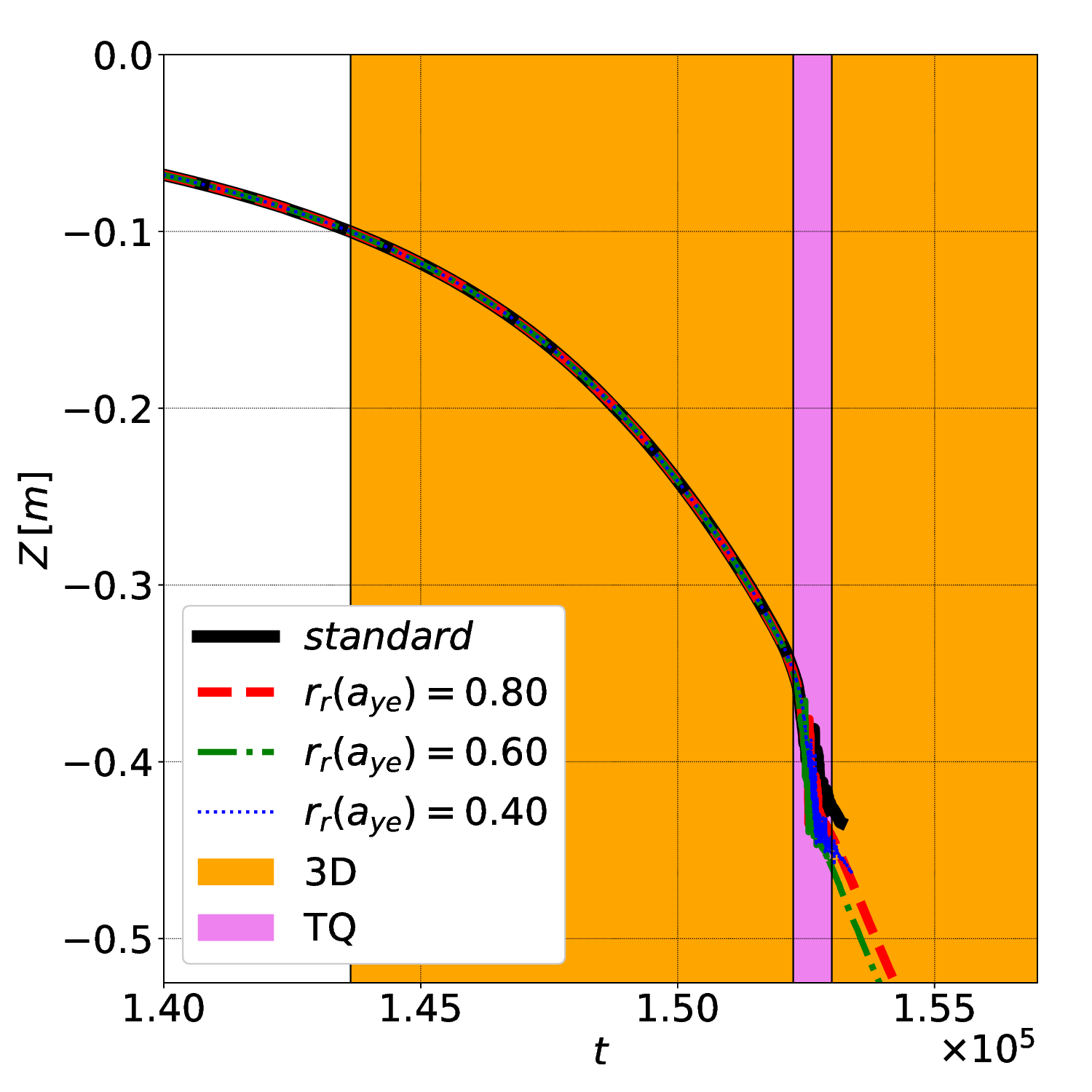}
        \caption{\label{fig:sfig08a}}
    \end{subfigure}%
    \begin{subfigure}{.5\textwidth}
        \centering
        \includegraphics[width=.9\linewidth]{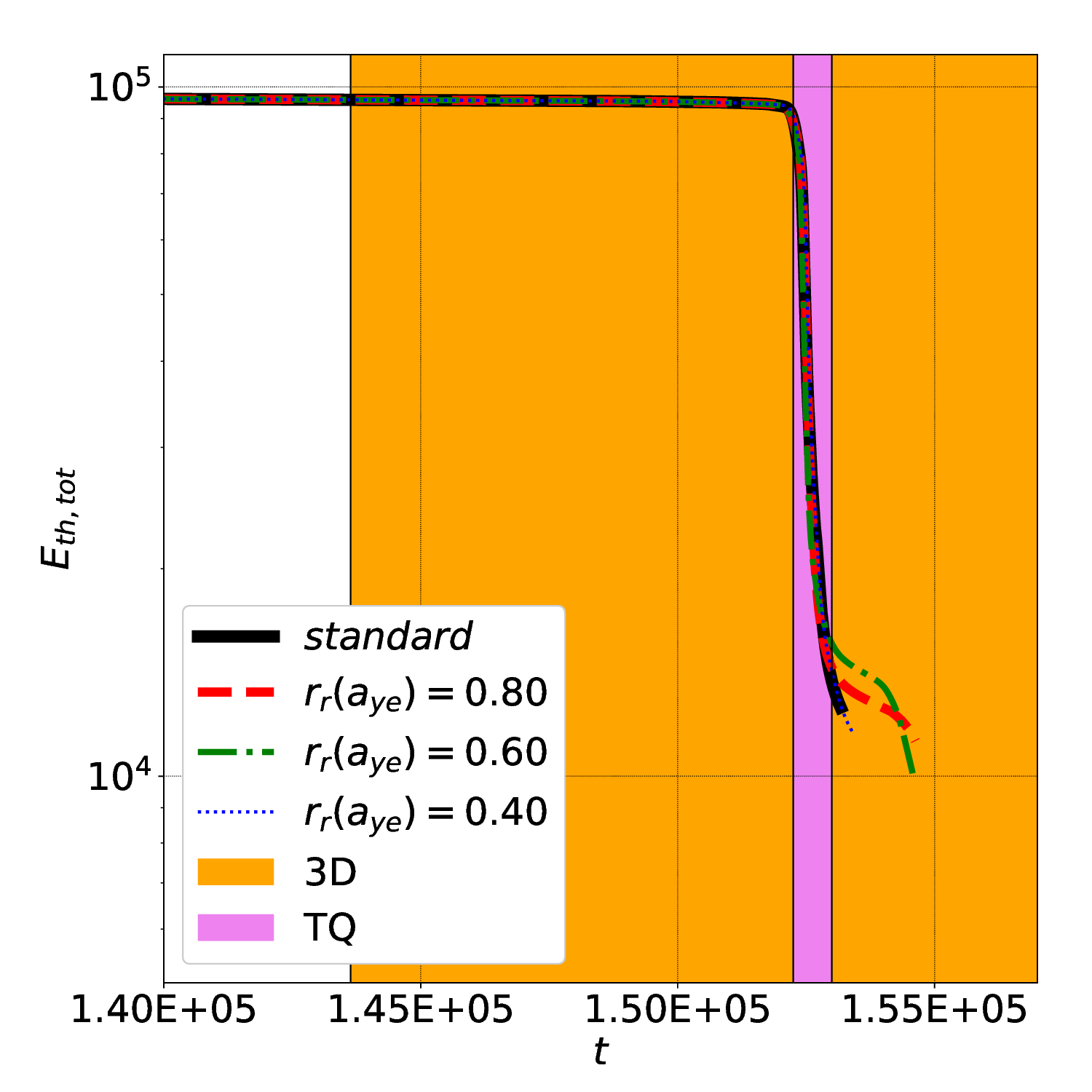}
        \caption{\label{fig:sfig08b}}
    \end{subfigure}
    \caption{\footnotesize\label{fig:fig08}Time evolution of the $Z$ position of the magnetic axis (figure~\ref{fig:sfig08a}) and the total thermal energy of the plasma (figure~\ref{fig:sfig08b}) for the simulations adopting $n_F=19$. Different line styles (with different colors in the online version) represent different rates of retained singular values for the compression, as shown by the legend. The highlighted vertical orange region locates the 3D phase, while the TQ phase is depicted with a vertical violet band, and its duration coincides with the one of the steep drop in the total thermal energy\normalsize}
\end{figure}

\begin{figure}[b!]
    \centering
    \includegraphics[width=0.9\linewidth]{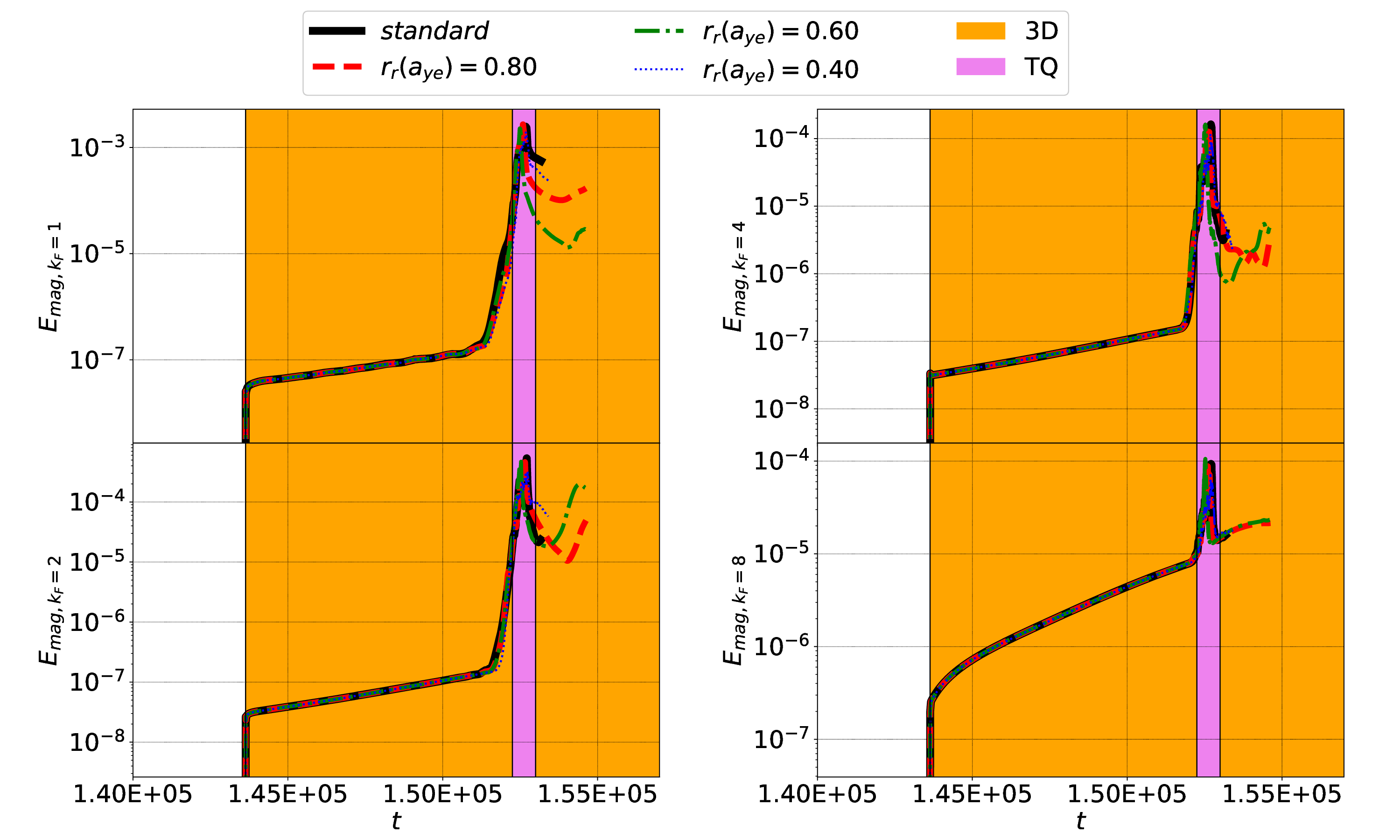}
    \caption{\footnotesize\label{fig:fig09}Time evolution of the $k_F = 1$, $2$, $4$, and $8$ of the magnetic energy for the VDE test adopting $n_F=19$. In the online color version, different colors depict different choices for the compression of $M^{ye}$, as shown in the legend at the top. The colored regions are the same as figure~\ref{fig:fig08}}
\end{figure}

Figures~\ref{fig:fig08} and~\ref{fig:fig09} show the time traces of the vertical position of the magnetic axis, the thermal energy content of the plasma, and the magnetic energies for selected harmonics when adopting $n_F=19$. The curves (colored in the online version) are obtained from uncompressed (indicated by standard) or compressed $M^{ye}$. The time considered in such plots spans from slightly before starting the 3D phase and up and beyond the TQ phase. In particular, figure~\ref{fig:fig08} shows that the results for the $Z$--coordinate of the magnetic axis and the total thermal energy obtained without compression are almost exactly matched by the ones obtained compressing the matrix $M^{ye}$, up to the rapid drop in the total thermal energy, happening during the TQ phase. Indeed, during the highly non-linear TQ, the $Z$ position of the magnetic axis shown in figure~\ref{fig:sfig08a} becomes less regular. In this phase, the compressed simulations cannot accurately describe the evolution anymore. Figure~\ref{fig:sfig08b} confirms a similar observation in terms of the thermal energy content of the plasma: cases with compressed $M^{ye}$ reproduce the dynamics of the original case up to the TQ phase, during which the accuracy has deteriorated. For the same cases, Figure~\ref{fig:fig09} compares the evolution of magnetic energies of selected toroidal components. Also here, the initially excellent agreement deteriorates during the TQ. It is worth noting, that the agreement in the higher harmonics is better than in the low harmonics.

The errors occurring during the TQ when adopting compression are likely due to the development of strong dynamical instabilities, usually able to break the magnetic confinement surfaces (see e.g. \cite{artola2023modelling}). Furthermore, through numerical experiments, we found that the plasma evolution appears particularly sensitive to any possible asymmetry in the conducting structures around the walls, during the TQ. Addressing such loss of accuracy during the evolution at advanced times, when strong dynamics are active requires careful treatment of the compression method from the point of view of the modeled geometry. Of course, such a task is of interest for future developments and deserves detailed additional explorations.

Moreover, understanding the reason why, compressing only $M^{ye}$ in VDE simulations produced accurate results while compressing $M^{ey}$, as well, produced inaccurate results could be related to the particular nature of the instability itself but requires further investigation and goes beyond the scope of the present work.

\section{Conclusion and Outlook}
\label{sec:Conc}

This article described compression methods for large dense matrices that occur in the coupling of JOREK to STARWALL or CARIDDI, explained the implementation in the parallel code, and verified its correctness. The two matrices $M^{ey}$ and $M^{ye}$ that describe the mutual interaction between plasma and conducting structures are addressed, which typically dominate the memory consumption. The tests focus on two representative cases with moderate resolutions: in tearing mode tests, good results were obtained, while the more involved vertical displacement events
do not show reliable behavior with compression, as discussed in detail in the article. Testing the limits of the method was an essential goal of this work.

In detail, in the simulation of a slowly growing TM instability, the axisymmetric component is kept fixed such that the response matrices only affect the non-axisymmetric components of the plasma evolution. Here, a strong reduction of the memory consumption can be obtained and an increasing compressibility is seen for both higher wall or plasma resolutions. 

The simulation of a 3D VDE behaves differently. Only the $M^{ye}$ matrix could be compressed, while $M^{ey}$ compression was unsuccessful. Furthermore, even at moderate compression levels, the initially excellent agreement deteriorates during the highly non-linear thermal quench phase. This VDE problem is thought to be a result of only considering the SVD spectrum for the compression, while the highly violent plasma dynamics might drive certain modes even though they are associated with smaller singular values, such that they shouldn’t have been neglected. Overcoming
this limitation will be an important step for the general applicability of the method, but requires additional analysis of the plasma side to complete the picture such that it has to be left for future work. Nevertheless, the method is already applicable to relevant cases with good compression results as demonstrated for the tearing mode example such
that the development has substantial practical benefits in its present form.

The overall test results are considered promising and will be the basis for future large--scale production applications that go beyond the scope of this article.
 
Presently observed limitations provide insights regarding the future directions of this work. At first, a plain SVD application limits the compression by the smallest matrix dimension, as understandable from equation (\ref{rr}) and visible from figure~\ref{fig:fig06}. The largest dimension of the matrices is clearly given by the DoF in the considered device's wall and grows significantly when increasing the resolution. Adopting more complex methods (see~\cite{cau2022fast} and references therein for some examples) might allow higher compression even when adopting more detailed wall geometries. The limitations in the case of strong dynamical instabilities suggest that geometrical aspects should be taken into account in future work. Finally, the JOREK-STARWALL and JOREK-CARIDDI couplings are now restricted to reduced MHD and do not account for halo currents flowing directly between conducting structures and the plasma domain. A full MHD treatment presently under development will give rise to new response matrices that can be compressed with the techniques described in the present paper and their successors.

Despite the limitations mentioned above, the novel implementation of the response matrices compression technique via SVD presented in this manuscript provided useful results that are already applicable while paving the way toward further improvements.

\section*{Acknowledgments}

This work has been carried out within the framework of the EUROfusion Consortium, funded by the European Union via the Euratom Research and Training Programme (Grant Agreement No 101052200 — EUROfusion). Views and opinions expressed are however those of the author(s) only and do not necessarily reflect those of the European Union or the European Commission. Neither the European Union nor the European Commission can be held responsible for them. Some of this work was carried out on the Marconi-Fusion supercomputer operated by CINECA. Code development and testing of the presented work have also been performed on the MareNostrum 4 cluster at the BSC, through the allocation of the Computer Applications in Science and Engineering (CASE) Department.

\section*{Data Availability}

The data that support the findings of this study are available from the corresponding author upon reasonable request.

\bibliography{bibliography}

\end{document}